# The Price of Quietness:

# How a Pandemic Affects City Dwellers' Response to Road Traffic Noise

Yao-pei Wang, Chinese Academy of Housing and Real Estate, Zhejiang University of Technology [1]

Yong Tu, Department of Real Estate, School of Business, National University of Singapore[2]

Yi Fan, Department of Real Estate, School of Business, National University of Singapore[3]

[1] Chinese Academy of Housing and Real Estate, Zhejiang University of Technology, Hangzhou, China. E-mail: wangyaopei123@163.com. ORCID ID:0000-0002-6729-7780.

[2] Department of Real Estate, School of Business, National University of Singapore, 15 Kent Ridge Drive, Singapore 119245, Singapore. Corresponding author and email: biztuy@nus.edu.sg. ORCID ID: 0000-0002-4581-3676.

[3] Department of Real Estate, School of Business, National University of Singapore, 15 Kent Ridge Drive, Singapore 119245, Singapore. Corresponding author and email: yi.fan@nus.edu.sg. ORCID ID: 0000-0001-6530-8282.

We acknowledge financial support from the National University of Singapore Academic Research Fund R-297-000-144-115 and Singapore Ministry of Education Social Science Research Thematic Grant R-311-000-030-119.

# The Price of Quietness:

# How a Pandemic Affects City Dwellers' Response to Road Traffic Noise

## Abstract

Using the outbreak of COVID-19 in Singapore as a quasi-natural experiment, we investigate tenants' changing responses to road traffic noise in the rental housing market, using 46,980 transaction records between 2006 and 2022. Our difference-in-differences estimates show that road traffic noise decreases housing rents by 3.8% immediately after the pandemic outbreak and further declines by 12.7% in the subsequent year—equivalent to 186.7 US dollars per month. The results are robust to parallel trend analysis, permutation placebo tests, and tests using alternative distance thresholds or distance to the nearest main road. Then, we adopt a machine learning text analysis of 10,425 rental housing advertisements, showing that tenants' preference for quietness increases by approximately 10% from 2019 into 2020. The new work-from-home business model and rising traffic from delivery services can explain for this pattern. To the best of our knowledge, this is the first paper using a large volume of transaction records to quantify city dwellers' willingness to pay for quietness in the COVID-19 context. Our results have policy implications for other nations and post-pandemic era on the interaction among urban planning, transport networks, and human settlements, and shed light on the pathway to achieve sustainable development goals.

## 1. Introduction

Cities are the hubs of human and economic development. One of the key Sustainable Development Goals (SDGs) proposed by the United Nations is to promote sustainable cities and communities, where sustainable transport plays a significant role to achieve the goal. In the transport literature, one inevitable byproduct is noise pollution which has been found to have negative health effects on people (Sygna *et al*., 2014; Pujol *et al*., 2014; Fan & Weinhold, 2022). It also directs urban residents to quieter homes, thereby causing value appreciation for properties with low exposure to traffic noise (Kuehnel & Moeckel, 2020; Diao *et al*., 2016; Fan *et al.*, 2021). The unexpected outbreak of COVID-19 has increased the sensitivities of city dwellers to urban noise in line of government policies such as social distancing and consequently, has brought a wide range of behavioral changes in social networks (Bavel *et al.*, 2020) and public health (Betsch *et al.*, 2020). It has also brought new challenges for the development of sustainable cities and transportation (Khalil & Fatmi, 2022; Zhang *et al.*, 2023).

Using the spread of the pandemic in a city-state of Singapore as a quasi-natural experiment, together with its compact urban form and developed transport networks, we evaluate city dwellers' changing willingness to pay for quietness. People are expected to be more exposed to a noisier environment during the pandemic and even afterwards, with more flexible working arrangements such as working from home (WFH) and modern communicational tools. Some employees may have to be exposed to road traffic noise that otherwise can be avoided by working in an office. Meanwhile, the pandemic unprecedentedly boosted e-commerce, as various policy responses to COVID-19 limited in-person shopping, causing people to turn to online shopping and accelerating the growth of the delivery industry. More delivery vehicles took to the roads, creating additional traffic congestion and noise as well as directing traffic towards homes. Taking these two effects into account, we aim to investigate how an unexpceted outbreak of the pandemic changes city dwellers' behavioral responses to road traffic noise and willingness to pay for quietness.

Using 46,980 housing transactions between 1 July 2006 and 31 March 2022 in the public open rental housing market obtained from the largest real estate data provider in Singapore, we set up a difference-in-differences (DiD) approach to empirically identify the causal effect of COVID-19 on changing behaviors toward road traffic noise. We first empirically identify the treatment group to include housing units within 100 meters of main roads, using the method of continuous threshold zones (Diao *et al*., 2016). The classification of treatment group is consistent with the literature (Fan *et al*., 2021; Lee & Pang, 2022). Then we estimate the additional rent premium charged for quiet accommodation after the pandemic, as a measure of city dwellers' willingness to pay for quietness. We find that for housing units with lower traffic noise exposure, tenants were willing to pay an additional rent premium of 3.8% in the first year after the outbreak of COVID-19. The willingness to pay more than trippled to 12.7% in the subsequent year. It is equivalent to a rise of 186.7 US dollars per month. Our results survive a battery of robustness checks.

To examine possible channels, we apply a machine learning text analysis method to 10,425 rental housing advertisements during the sample period. We find that tenants' preference for quietness increases by approximately 10% from 2019 to 2020, reaching 1/3 of the overall advertisements. Possible reasons for the increasing preference lie in the new WFH business model and rising traffic from delivery services.

To the best of our knowledge, this is the first paper using a large volume of transaction records to quantify city dwellers' willingness to pay for quiet home environment in the context of a pandemic. We contribute to the traffic noise literature (Wilhelmsson, 2000; Baranzini & Ramirez, 2005; Ossokina & Verweij, 2015; Diao *et al*., 2016; Kuehnel & Moeckel, 2020; Fan *et al.*, 2021) and to the debate on human behaviors in a post-COVID-19 environment (Audrin *et al.*, 2022; Bracarense & de Oliveira, 2021; Rosenthal *et al*., 2022; Mouratidis & Yiannakou, 2022; Mouratidis, 2022). We attribute the behavioral shifts to the e-commerce and WFH boost resulting from the social distancing preventive measures in place during the pandemic. As e-commerce and WFH continue gaining traction in digitally mature markets, such behavioral changes can pose challenges to post-pandemic regulations on urban traffic noise, noise pollution control and urban design in promoting health and well-being.

Our research also adds to the burgeoning literature on sustainability (Beck & Ferasso, 2023; Wang & Sun, 2022) and has policy implications for nations beyond Singapore. "Developing sustainable cities and communities" is a major goal under the SDGs; sustainable transport plays a significant role in achieving this goal. Kar *et al*. (2022) find that the pandemic has severely influenced public transit, thereby worsening accessibility for communities with multiple social vulnerabilities. We contribute to this stream of literature by identifying and quantifying the increasing willingness to pay for quiet home environments among urban residents. Our results call for attentions of policymakers to endogenize city dwellers' changing responses to urban noise into the development of sustainable transport networks. The findings also emphasize on the interaction between urban planning and the quality of human settlements, shedding light on the pathway to achieve sustainable development goals.

## 2. Institutional background

Public housing represents 80% of the housing stock in Singapore. The high-density, housing-estate-based residential urban design physically exposes some buildings to more traffic noise by being closer to the main roads, which also blocks road traffic noise for others and renders them less exposed to traffic noise (Figure 1). In Singapore, the average level of traffic noise in some public housing neighbourhoods was 65 dB(A) in 2004 (Chui *et al.*, 2004) and increased to 71.8 dB(A) in 2017 (Ng & Xi, 2017)[1]. For comparison, the World Health Organization

[1] The Singapore government does not have a regular nationwide practice of measuring urban noise. Data on the noise levels from several specific case studies are not publicly available. Instead, we use two alternative proxy

recommended that community noise should be maintained at less than 30 dB(A) in bedrooms for a good night's sleep and less than 35 dB(A) in classrooms to ensure a good teaching and learning environment.

The majority of Singapore's public housing dwellers are public homeowners. Tenancy agreements are typically renewed annually at a market rate, thus promptly reflecting the changes in a renter's willingness to pay for a housing unit (Tu & Bao, 2009). Melser (2020) suggests that rental housing transaction data are more suitable for road traffic research than housing sales data because rents reflect the use values of the house while prices internalize the investment values. Traffic noise affects housing decisions mainly through consumption; thus, rental housing transactions constitute desirable data for this study. Besides Singapore, rental housing is popular in other Asian countries, with market share ranging from 20% in China (Huang *et al.*, 2020) to around 50% in Korean (Kyung-Hwan & Miseon, 2016).This paper investigates the impacts of traffic noise from main roads on public housing. There are four types of roads in Singapore (Table A1 in Appendix). Local and collector roads have lower speed limits and fewer lanes, thus generating less traffic noise than main roads and expressways. The main roads are longer and have a wider coverage of residential areas than expressways, because expressways connect locations and main roads connect housing neighbourhoods. Figure 1 illustrates the interaction of main roads with residential buildings. Some sections of the expressways in Singapore have sound barriers, which may cause estimation bias (Sygna *et al.*, 2014). Section 4 reports how we isolate the impacts of expressway, train (MRT), and airport traffic noises from the impacts of main road traffic noise on a housing unit.

Singapore confirmed its first COVID-19 case in January 2020, and that year witnessed a two-month circuit-breaker lockdown followed by a three-phase approach to gradually resume

---

variables that are well-established in the literature to measure traffic noise, which will be introduced in detail in Section 4.

social and economic activities. Between January 2021 and March 2022, Singaporeans survived several waves of surges in COVID-19 cases, followed by a gradual policy relaxation moving toward living with COVID-19. During both periods, Singapore's e-commerce remained resilient and expanded rapidly, whereas the population share of WFH fell (Table A2 in Appendix).

## 3. Literature review and research hypotheses

### 3.1 The impact of Traffic noise on Housing markets

Noise pollution is one of the most important environmental risks to health and has triggered growing concerns among policymakers and the public (WHO, 2018), and traffic noise is the biggest culprit of such polluting sources (Das *et al.*, 2019). For instance, it may disturb sleep (Sygna *et al*., 2014), have cardiovascular effects (Babisch, 2011), result in poorer work and school performance (Pujol *et al*., 2014) and negatively affect mental health (Ma *et al*., 2020).

In the traffic noise literature, the housing market is often used to study the socioeconomic consequences of traffic noise (Wilhelmsson, 2000 & 2005; Baranzini & Ramirez, 2005; Ossokina & Verweij, 2015; Diao *et al*., 2016; Kuehnel & Moeckel, 2020; Yao *et al.*, 2021; Fan *et al.*, 2021). For example, traffic noise negatively influences the housing unit prices (Diao *et al.*, 2016; Kuehnel & Moeckel, 2020); increases noise complaints from urban residents (Fan *et al.*, 2021); and causes heterogeneous housing prices because of variations in preferences for quiet housing units (Von Graevenitz, 2018). However, to the best of our knowledge, this is the first paper using a large volume of transaction records to quantify the price of quietness that people are willing to pay in the COVID-19 context. Our research not only estimates the socioeconomic consequences of traffic noise, but also uncovers the changes in city dwellers' behavior during COVID-19 in housing rental markets.

### 3.2 COVID-19 and urban behavior changes

Social distancing was a major preventative measure in containing the pandemic, directly curtailing personal mobility and outdoor activities (Bracarense & de Oliveira, 2021; Khalil & Fatmi, 2022). Most research studied how the resulting travel behavior changes influence the urban environment, such as air quality improvement (Benchrif *et al*., 2021; Latif *et al*., 2021) and changes in the acoustic environments (Rumpler *et al*., 2020; Basu *et al*., 2021).

To facilitate social distancing, people were strongly encouraged, or enforced, to remain at home. Employers also adopted flexible working arrangements to accommodate such policies. These changes triggered behavioral responses in their housing preferences. Psychology studies have shown that extended periods of home confinement can change residents' sensitivities toward their environment (Lee & Jeong, 2021; Marroquín *et al*., 2020). In the context of the COVID-19 pandemic, it was also found that the compactness of neighbourhoods was correlated with a lower level of well-being using Norwegian data (Mouratidis, 2022) and living closer to public transportation facilities can increase residents' level of anxiety using Greek data (Mouratidis & Yiannakou, 2022). It is also find that the city residents value more on cultural ecosystem services (CES) of urban and peri-urban forests (Beckmann-Wübbelt *et al.*, 2021).

Economists claim that the pandemic has had an extensive influence on the urban retail sector and that retailers should reconfigure towards agile retailing in the time of the digital economy (Pantano *et al*., 2020). The lockdown policy pushed consumers towards online shopping, as social distancing measures mandated that physical stores had to shut down temporarily during lockdowns. This accelerated the development of e-commerce (Abdelrhim & Elsayed, 2020). It also resulted in a shift in preferences towards e-commerce (Hu *et al.*, 2020), and academics expect such trends to become the "new normal".

A few relevant studies found that COVID-19 reduced the value of density and flattened rent gradients in American cities that rely heavily on rapid rail transit but not in car-oriented cities (Rosenthal *et al.*, 2022). They interpreted these changes as indicating that residents devalue the accessibility to CBD areas (Liu & Su, 2021). Several literature measure the outdoor urban noise level during lockdown period (Rumpler *et al.*, 2020; Basu *et al.*, 2021;Lee & Jeong, 2021). Lee & Jeong (2021) find that although the total level of urban noise declined during lockdown period, the neighborhood noise complaints increased. They explain that resident may receive more noise in home than in office. We make an original contribution to the above literature by adopting a DiD approach to identify the causal effect of COVID-19 on the changes in human behaviors regarding road traffic noise using housing market data and by interpreting our findings using the boosting of e-commerce and WFH. We focus on how city dwellers' change their willness to pay for quietness in post-pandemic era rather than the change of urban noise level.

### 3.3 Research hypotheses and Channels

We use a tenant's willingness to pay an additional rent premium for a housing unit with lower traffic noise exposure to measure the behavior change. Thus, the first hypothesis is that COVID-19 may cause urban residents to be willing to pay an additional rent premium for a housing unit with lower traffic noise exposure.

Social distancing reduced in-person contact and subsequently induced changes in residents' behaviors concerning road traffic noise through two channels: WFH and e-commerce. For example, adherence to social distancing shifted more employees to remote work in 2020,[2] and more people avoided public events and replaced outdoor activities with in-home activities (Leger, 2020). When people spent more time at home during COVID-19, not only was their travel demand reduced (Bracarense & de Oliveira, 2021), but this might also have rendered them less tolerant of road traffic noise, which they previously avoided by working in offices.[3]

---

[2] The government policies can be retrieved at https://stats.mom.gov.sg.

[3] Singaporean residents' noise complaints related to their daily activities increased threefold, from 3,600 cases in 2019 to 11,400 cases in 2020; 49% employees worked from home in 2020. Source: https://www.mnd.gov.sg and https://stats.mom.gov.sg

Not every home has a quiet workplace, which may have enhanced people's preferences for quiet homes. Meanwhile, social distancing has accelerated the pre-existing trend of e-commerce (Kim, 2020). For example, the fear of the pandemic improved consumers' perceptions of the social economic benefits of e-commerce (Tran, 2021). Faster delivery has become a new normal that has increased traffic flow toward homes and created burdens for cities with more congestion, higher traffic volumes, more noise and air pollution (Ranieri *et al.*, 2018; World Economic Forum, 2020). The benefits from the reduced demand for travel from shopping might be cancelled out by the increased number and frequency of delivery vehicles if the last-mile deliveries are not well-managed. More traffic noise would naturally reinforce the already existing preference for quiet homes.

To support our justification, we use two proxy variables, which are well-established in the literature, to measure traffic noise. First, we collected 6,093 and 4,332 rental housing advertisements in 2019 and 2020 respectively from a Singaporean online housing portal (SGRoom.com) and use people's subjective perception of traffic noise as an indicator of noise level (Fan *et al.*, 2021; Friedt & Cohen, 2021). A machine learning text analysis shows that the rental housing advertisements mentioned 'quiet' increased by approximately 10% from 2019 to 2020, reaching 1/3 of the overall advertisements. Of the top 20 most frequently used words in a rental housing advertisement, 'quiet' rose from eleventh (2019) to eighth (2020). We therefore conjecture that during COVID-19, residents valued quiet homes more, giving rise to an additional rent premium. Second, we measure traffic noise by using the distance to transportation facilities (Diao *et al.*, 2016; Von Graevenitz, 2018; Pinsonnault-Skvarenina *et al.*, 2021) and apply a Difference-in-Differences method to estimate the casual effect of COVID-19 pandemic on changes of city dwellers' willingness to pay for quietness. The detailed empirical strategy is discussed in following section.

The second hypothesis is that the additional rent premium for a housing unit with low traffic noise exposure is increasing from 2020 to 2021. First, in 2021, the population share of WFH decreased, indicating that commuting trips should have risen. E-commerce remained strong with an annual revenue growth rate of 37.5% (Table A2 in Appendix). A smooth online customer experience relies heavily on the convenience and speed of receiving an order, thus an e-commerce boom turbocharges the goods delivery industry directly. Hence, in 2021, Singaporeans had more traffic congestion and noise than in 2020 arising from delivery services. Additionally, housing turnover incurs housing search and transaction costs. People would not move if the benefits of moving could not cover the search and transaction costs. The year of 2021 witnessed several coronavirus waves, which likely rise city dwellers' willingness to pay for quietness.

## 4. Research design and data collection

### 4.1 Identification of the treatment and control groups

The hypotheses are tested using a DiD model. We adopt a semi-log hedonic pricing model to examine how the Covid-19 pandemic changes city dwellers' willingness to pay for quietness. We rely on two dimensions of variation in this model: distance to main road (treatment vs control group), and before and after the outbreak of COVID-19 pandemic.The first step is to estimate a distance threshold (Equation 1), which is used to classify housing units into a treatment or a control group. Housing units in the treatment group are significantly influenced by road traffic noise. We use a semi-log hedonic regression, with logarithm of rent as the outcome variable following the literature (Rosen, 1974). Taking the logarithm form reduces estimation biases (Sopranzetti, 2015) and facilitates interpretations using the percentage term (Clapp & Salavei, 2010). We choose a 50-meter increment to construct seven threshold zones based on previous studies (Autor *et al.*, 2014; Diao *et al.*, 2016; Diamond & McQuade, 2019). Specifically, if the distance of a building to the nearest main road is less than or equal to 50 meters, it falls into threshold zone 1, indicated by 0–50 m. Threshold zones 2–6 are 50–100 m, 100–150 m, 150–200 m, 200–250 m, and 250–300 m respectively. The omitted group includes rental transactions located more than 300 m away from the main roads (zone 0).

Buildings closer to the main roads were exposed to more road traffic noise but were more accessible. In the absence of traffic noise, we would expect the accessibility of the buildings from threshold zone 1 to zone 6 to decrease monotonically. In other words, the values of $\alpha_1$ to $\alpha_6$ are positive but decrease monotonically. Given the existence of road traffic noise, which has a negative impact on the values of $\alpha_1$ to $\alpha_6$, we may not necessarily see a linearly decreasing trend from $\alpha_1$ to $\alpha_6$ because of the opposite impacts of road traffic noise and accessibility on rent. However, the turning point from a significant and negative $\alpha_{i-1}$ to a (insignificant and) positive $\alpha_i$ indicates that main road traffic noise on rent no longer affects rent when we move from threshold zone *i-1* to threshold zone *i*, thereby giving rise to the distance threshold. Subsequently, if a building's distance to the nearest main road is less than the distance threshold, the housing transactions in this building are classified into the treatment group. Otherwise, they are classified in the control group. A hedonic housing rent model is used to estimate the housing rent premium (or discount) of each threshold zone.

$$lnRent_{ijdt} = \alpha_0 + \sum_{i=1}^{6} \alpha_i group_i + \gamma X_{ijdt} + \tau_j + \pi_d + \mu_t + \varepsilon_{ijdt} \quad (1)$$

Here, $lnRent_{ijdt}$ represents the natural logarithm of monthly housing rent in threshold zone *i*, of housing type *j*, in district *d*, and year-month *t*. $X_{ijdt}$ is a vector of control variables including housing unit characteristics and geographic information. The housing unit characteristics include housing unit size (Size), floor level (Floor), and the age of a housing unit (Age) which are classical hedonic variables used in the literature (Diao *et al.*, 2016; Fan *et al.*, 2021). Since location is always an important factor, if not the most, for monthly rent (Jones & Simmons, 1990; Verbrugge *et al.*, 2017), we also include a bundle of geographic information of housing building: Distance to nearest expressway, MRT station, primary school, Central Business

District (CBD), and bus stop (Baranzini & Ramirez, 2005; Bagheri & Shaykh-Baygloo, 2021). The definitions of variables are summarized in Appendix Table A3.

To reduce the omitted variable bias brought by unobservable factors, we include three additional sets of fixed effects: housing type-fixed effects, $\tau_j$ which control for unobserved characteristics of housing unit across different types (details in Appendix Table A3); district fixed effects, $\pi_d$ which are used to control for unobserved spatial features at different planning areas, and year-month fixed effects, $\mu_t$ which account for price patterns across different years and seasonality within the same year. Error term $\varepsilon$ is clustered by each residential building to capture autocorrelations.

**4.2 The DiD model**

The second step is to construct a DiD model and test our research hypotheses (Equation 2). Our model captures spatial variation, which reflects the impact of road traffic noise on housing rent, and temporal variation, which reflects the impact of COVID-19 on housing rent. Thus, our DiD model calibrates the causal relationship between the COVID-19 shock and residents' behavior changes in response to road traffic noise.

$$lnRent_{ijdt} = \beta_0 + \beta_1 DID_{it} + \beta_2 Treatment\ group_i + \beta_3 Covid_19_t + \gamma X_{ijdt} + \tau_j + \pi_d + \mu_t + \varepsilon_{ijdt} \quad (2)$$

Where $lnRent_{ijdt}$ indicates the natural logarithm of monthly housing rents in threshold zone *i*, of housing type *j*, in district *d*, and in year-month *t*. $Treatment\ group_i$ is a dummy variable, wherein 1 indicates a housing unit $i$ in the treatment group and 0 otherwise. $Covid_19_t$ is a dummy variable with 1 indicating that a housing unit was transacted post COVID-19 (1 January 2020) and 0 otherwise. $DID_{it}$ is the interaction term of $Road\,noise_i$ and $Covid_19_t$. Definitions of other variables are the same as in Equation (1). To examine the potential presence of multicollinearity in our model, we conduct a Variance Inflation Factor (VIF) test. The VIF value is 1.83, which is far below the multicollinearity threshold of 10 (Salmeron *et al.*, 2018) and close to 1, indicating the absence of any milticollinearity issue (James *et al.*, 2013).

Coefficient $\beta_1$ measures the additional housing rent discount for a housing unit with high traffic exposure that was transacted post-COVID-19. It captures average treatment effect from pandemic on the rent of housing units which suffer road traffic noise problem. Given that we are using the logarithm of rent, the coefficient $\beta_1$ directly quantifies the percentage discount that the treatment group experiences during the pandemic period. 

To test hypothesis 2, we divide the post-pandemic period into two sub-periods: 2020 and between 1 January 2021 and 31 March 2022. Hypothesis 2 holds if $\beta_1$ is (significantly) more negative in the second sub-period than in the first.

We undertake four validity tests. The parallel trend test (Jacobson *et al*., 1993) posits that in the absence of treatment, the difference between the treatment and control groups in a DiD model is constant over time. The permutation placebo tests (Lu & Anderson, 2015) hold that no other shocks can cause significant differences between the treatment and control groups, and the robustness test shows that the findings are not sensitive to the identification of treatment and control groups (Von Graevenitz, 2018). Finally, the heterogeneity test is conducted to capture variations in the additional rent premium across noise levels and dwelling sizes.

**4.3 Data collection**

Rental housing transactions between 1 July 2006 and 31 March 2022 in the Singapore public open rental housing market were obtained from the *99 Group*, which is a real estate technology firm and the largest real estate data provider in Singapore. The high homogeneity in public housing reduces the impact of unobserved factors on rental values.The dataset records monthly housing rent, contract date, housing attributes and address. We used the ArcGIS tool to measure the distance from a building to the nearest main road and to construct other spatial variables. We excluded housing transactions that can be exposed to other types of traffic noise (e.g., located in the vicinity of airports) and obtained 87,069 transactions. The exclusion rules are consistent with existing literature: the impact of traffic noise from an expressway on housing prices disappears after 300 meters (Pinsonnault-Skvarenina *et al*., 2021), the impact of railway noise on housing prices disappears after 400 meters (Diao *et al*., 2016) and the impact of airport noise on housing prices disappears after 3,000 meters (Von Graevenitz, 2018). We restrict control group to be within 500 meters and the working dataset includes 46,980 transactions. All variables are defined in Appendix Table A3 and their summary statistics are presented in Table 1.

## 5. Empirical results

Section 5.1 presents the findings using Equation (1), and the treatment and control groups are identified. Section 5.2 presents the main findings using Equation (2). Section 5.3 presents the results of the parallel trend tests, and Section 5.4 presents the permutation placebo tests, robustness and heterogeneity tests.

### 5.1 Identification of the treatment and control groups

Equation 1 is applied to the full sample (between 1 January 2006 and 31 March 2022) and the pre-pandemic subsample (between 1 January 2006 and 31 December 2019) separately (Table A4 in Appendix). The results in both columns suggest that the distance threshold is 100 meters (see Section 4.1 for an explanation). This is consistent with the findings of Fan *et al.*'s (2021) Singapore-based road traffic noise study.

Using the 100-meter distance threshold estimated in Table 1, the residential buildings in the sample are classified into the treatment and control groups. The transactions in the treatment group units are within 100 meters of a main road, consisting of 13,597 samples. To ensure that transactions in the control group have hedonic features similar to those in the treatment group, the transactions in the control group are limited to housing units located between 100 and 500 meters from a main road, which results in 33,384 samples. In total, we have 46,980 transactions. The 500-meter cut-off was used by Diao *et al*. (2016), Fan *et al*. (2021) and Lee & Pang (2022), which are Singapore-based traffic noise studies. The summary statistics for treatment and control groups are given in Table 1 and the spatial interactions of the residential buildings with the main road lines are illustrated in Figure 1.

[Place Table 1 about here]

[Place Figure 1 about here]

In summary, the average unit sizes are 91.66 $m^2$ and 99.26 $m^2$ for the treatment group and control group, respectively, and the average floor levels are 7.80 and 8.14, respectively. Thus, the property attributes of the two groups of housing units are not significantly different. The average distance from a residential building to the nearest main road was 61.44 meters in the treatment group and 272.60 meters in the control group.

**5.2 Hypothesis tests**

Equation (2) is applied to the full sample and two sub-samples between January 2020 and December 2020 and between January 2021 and March 2022. Results are presented in Table 2.

[Place Table 2 about here]

The coefficients of *DID* across the four models are statistically significant with the expected signs. Housing units within 100 meters of a main road received an additional rent discount of 8.3% compared with housing units between 100-500 meters of a main road between January 2020 and March 2022. In other words, tenants were willing to pay an additional 8.3% rent premium for a housing unit with lower traffic exposure after the outbreak of the pandemic. In 2020, the additional rent premium was 3.8% and increased rapidly to 12.7% between January 2021 and March 2022. Both of the two estimates are statistically significant at a high 1% level.

The coefficients of the control variables are consistent with the literature. We find three significant results from control characteristics: First, larger and newer housing units and housing units on higher floors charge a higher rent. Second, tenants pay a rent premium for a housing unit closer to the Central Business District (CBD) or a train (MRT) station and pay a rent premium for quieter housing units. We also found that housing rents increased during the pandemic. During the investigation period, foreign workers, especially Malaysian workers who commuted to Singapore daily for work, were stuck in Singapore due to the pandemic-related border restrictions, which may have increased housing rents.

### 5.3 Parallel trend tests

To perform the parallel trend test, we designed seven time intervals, including: *more than 3 years before the pandemic* (i=-3); *3 to 2 years before the pandemic* (i=-2)*; 2 to 1 years before the pandemic* (i=-1)*; 1 to 0 years before the pandemic* (i=0)*; 0 to 1 year after the pandemic* (i=1)*; 1 to 2 year after the pandemic* (i=2) *and more than 2 years after the pandemic* (i=3). The parallel trend test is specified by Equation 3, which is applied to the full sample. The time interval of 'i=-3' is taken as the base. A negative $\alpha_i$ indicates a rent discount caused by road traffic noise. The results are presented in Table A5 in the Appendix and illustrated in Figure 2.

$$lnRent_{ijdt} = \sum_{i=-2}^{3} \alpha_i Treatment\ group_i \times relative_time_t + \beta Treatment\ group_i + \gamma X_{ijdt} + \tau_j + \pi_d + \mu_t + \varepsilon_{ijdt} \quad (3)$$

Here, $relative_time_t$, $t \epsilon \{-2,-1,0,1,2,3\}$ is a time dummy variable, with 1 indicating that a transaction was completed in the $t^{th}$ time interval and 0 otherwise. The cross term $Road\ noise_i \times relative_time_t$ indicates the impact of $Road\ noise$ on rent within the time interval.

[Place Figure 2 about here]

Figure 2 shows that before the pandemic started, the estimates of $\alpha_i, i = -2, -1\ and\ 0$ were insignificant with similar values (Table A5 in Appendix), indicating that housing rents in the treatment and control groups were subject to a parallel trend. After the pandemic, the estimates of $\alpha_i, i = +1, +2, and + 3$ were statistically significant and became larger over time.

The parallel trend assumption holds, meaning that the shock of COVID-19 enlarges the difference in the rent discounts caused by road traffic noise between the housing units in the treatment and control groups.

### 5.4 Permutation placebo tests, robustness and heterogeneity tests

The permutation placebo tests adopt the framework proposed by Lu & Anderson (2015) and used by Tang *et al*. (2020). It is used to examine whether the additional rent premium charged for quiet during COVID-19 might also be caused by other random factors. In this paper, we ran the baseline model (Model 1, Table 2) 1,000 times by randomly choosing the starting point of the pandemic. The outcomes are shown in Figures 3.1 and 3.2.

Figure 3.1 illustrates the distribution of the 1,000 estimated DiD coefficients. The mean of the 1,000 estimated DiD coefficients, indicated by the vertical line in Figure 3.1, is close to 0, implying that when we randomly select the pandemic starting date, the additional rent discount for a noisy housing unit during COVID-19 is very small compared with the DiD coefficient in Model 1 of Table 2, when the starting date is 1 January 2020. Figure 3.2 presents the scatter diagram of the p-values against the 1,000 estimated DiD coefficients. Most of the p-values are larger than 10% (the dots above the horizontal dotted line), meaning that most of the estimated DiD coefficients are not statistically significant. The p-values indicated by the dots below the horizontal dotted line are larger than 0.01, which is the corresponding estimate in the baseline analysis. Thus, the baseline estimation is adopted in this study. The permutation placebo tests that the additional rent discounts for noisy housing units (Table 2) are caused by the pandemic, starting from 1 January 2020.

[Place Figure 3.1 about here]

[Place Figure 3.2 about here]

Table 3 presents the robustness test. We reran the baseline model (Model 1, Table 2) by using 90 meters, 110 meters and a building's linear distance to the nearest main road to replace the 100-meter distance threshold and to rerun the baseline model (Table 3). All the results are significant and consistent with the findings in Table 2. In addition, we conduct a back-of-envelope calculation on the increase in WTP for a 1 dB(A) increment after the pandemic. Using distance as a proxy variable, the regression results in Table A6 in Appendix indicate that city dwellers are willing to pay an additional 0.352 SGD per meter to be further away from traffic noise in the post-pandemic era. The total WTP per meter for residents is 0.352 + 0.014 = 0.366 SGD. According to a Channel NewsAsia report, Singaporeans suffer approximately 100 dB(A) of road noise because of motocycles along[4]. Following the *Guidelines for community noise* given by WHO (Berglund et al., 1999), the sound level of normal speek is 50 dB(A). Therefore, we assume that the control group (distance>100 meter) receive an acceptable noise level of 50 dB. The average decrease of noise level is 0.5 dB(A) per meter when distance <= 100 meter: (100 dB(A)- 50 dB(A))/100 meter = 0.5 dB(A) per meter. Therefore, the average increase in WTP for a 1 dB(A) increment after the pandemic is 0.732 SGD ( $\frac{0.366\ SGD/meter}{0.5\ dB(A)/meter} = 0.732\ SGD/dB(A)$), around 0.55 USD/dB(A).

[Place Table 3 about here]

To identify the heterogeneous effects of traffic noise, we divide the full sample into six sub-samples using floor level, unit size and the distance to the nearest MRT station. In Singapore, housing units at lower floor levels or near an MRT station are likely to be in a noisier physical environment. Larger housing units are likely to have more variations of facing and thus suffer less from traffic noise. To test heterogeneity across different floor levels and housing types, we repeat the baseline model (Model 1 of Table 2) in sub-sample analyses and present the results in Table 4.

[Place Table 4 about here]

[4] https://mustsharenews.com/motorcycles-noise-level/

We found that in a noisier physical environment, tenants are significantly more willing to pay a higher rent premium for quieter homes during the pandemic, comparing the *DID* coefficients in Models 1 and 3 (a noisier environment) with those in Models 2 and 4 (a quieter environment) in Table 4. Comparing the *DID* coefficients in Models 5 and 6, we found that tenants renting a larger housing unit are more willing to pay for a higher rent premium for quieter homes during COVID-19. One interpretation is that tenants who rent larger units are likely to have a higher income and thus, hold an office job. Office workers are more likely to shift to remote work during the pandemic, which may have generated a higher demand for a quiet environment.

## 6. Discussion

### 6.1 External validity to nations beyond Singapore and the post-pandemic era

Literature has mixed evidence on the impact of pandemic on urban noise. Some studies show that lockdowns have reduced the overall urban noise level in developed countries, such as Sweden (Rumpler *et al.,* 2020), Ireland (Basu *et al.,* 2021), and the United Kingdom (Lee & Jeong, 2021). Others find that the neighborhood noise complaints increase during the pandemic (Lee & Jeong, 2021; Tong *et al.,* 2022). Because of the increasing WFH settings, city dwellers could be more sensitive to urban noise during and after the pandemic and exhibit heterogeneous patterns across different socioeconomic groups. The urban features of the city-state of Singapore, such as the dense city topology and ungated public housing residential buildings, provide valuable institutional settings to evaluate the additional premiums that city dwellers would like to pay for quietness and the changing patterns after the outbreak of the pandemic.

Our findings can be generalized to a broader context as e-commerce and work-from-home (WFH) practices become prevalent, with the ongoing expansion of digital marketplaces and e-commerce in the post-pandemic era. However, it is important to address the challenges associated with delivery services, particularly the impact of last-mile delivery on road traffic, through effective regulation and management (Allen *et al.*, 2018). Additionally, the WFH trend is not merely temporary during the pandemic, as evidenced by the doubling of remote workers every 15 years in the US prior to 2020 and the significant surge observed during the pandemic.Choudhury (2021) explored the benefits and concerns of remote working, calling for companies to embrace remote working (Kaushik & Guleria, 2020). As it becomes more prevalent, the socioeconomic consequences of remote working have become evident; for example, it has caused significant shifts in office lease revenues, occupancy and lease renewal rates (Gupta *et al.*, 2022). Given these trends, we expect preferences to lean towards quiet living spaces even beyond the pandemic and thus call for attention from policymakers on the sustainable city planning by incorporating this trend into account.

Specifically, this paper encourages policymakers and urban planners to have deeper consideration in how road networks can well interact with living spaces in urban development. Our study highlights that residential complexes should be built isolated from busy roads, as noise pollution from these roads negatively impacts the welfare of surrounding dwellers, worsening their standard of living. This is especially severe in urban areas with higher degree of flexible working arrangements and with more diverse online communicational tools. As populations in major cities increase, space constraints will only become tighter to accommodate more people, and thus greater importance should be placed on smart city planning. Previous literature has noted how green spaces, building design and spatial planning of residential complexes can potentially offset traffic noise pollution (Yuan *et al.*, 2019). There has also been research on how different road types affect noise pollution in nearby residential blocks (Fernandes *et al.*, 2020). We hope that our findings can add to this stream of literature by showing how the pandemic has resulted in a shock to preferences against noise pollution, changing the willingness to pay for a specific urban feature, quietness, over the longer term.

**6.2 Human settlements, transport Networks, and sustainability**

By 2030, nearly 60% of the world's population, approximately 5 billion people, will reside in urban areas. Rapid urbanization has brought about pressures on living environments and public health. Especially after the outbreak of the pandemic, there have been profound changes in the behavioral patterns of urban residents (Mahmud *et al.*, 2023; Rodriguez *et al.*, 2023). The challenge we face is how to establish sustainable cities and communities in the new normal. Sustainable transport is at the heart to make cities and human settlements inclusive and resilient. Singapore announced its "Green Plan 2030" in 2021, with one key pillar to develop sustainable living by promoting public transport and repurposing roads. Other nations have also set up similar sustainability plans, such as reducing private transport and encouraging the use of public or sharing vehicle alternatives: Many countries encourage car-sharing modes to decrease "Greenhouse Gas Emissions", for instance, the UK (Le Vine & Polak, 2019), Switzerland (Balac *et al.*, 2017) and Australia (Dowling & Kent, 2015). China practiced free-floating bike sharing, and this mode has a positive impact on urban air quality (Wang & Sun, 2022). In 2020, Luxembourg made the historical decision to introduce free public transport, the first such policy in the world. This is in addition to bike sharing movements gaining traction in metropolitan cities, with an expanding literature studying their effectiveness as part of a city's public transport network (Zhou *et al.*, 2020).

We hope to contribute to this movement by demonstrating the significance of endogenizing the changing request for human settlements from city dwellers into urban planning. Several European countries have been pioneers along this line, such as publishing monthly noise indexes to improve information disclosure to residents in Copenhagen, Denmark (Von Graevenitz, 2018); building noise production ceilings along highways in Poland (Szopińska *et al.*, 2022), or encouraging light vehicles for last-mile delivery in London, the UK (Allen *et al.*, 2018). Our findings call for more regulations and actions in nations aiming to achieve the sustainability goals.

## 7. Conclusion

This paper examines the negative externalities of road traffic noise in the post-pandemic era. We find that housing units with more exposure to road traffic noise have an additional rent discount of 8.3% and that tenants are willing to pay an additional rent premium for quieter housing units after the pandemic. We demonstrate that the policies implemented to keep social distance like WFH and digitalization after the outbreak of COVID-19 pandemic have enhanced people's requirement for quietness. We expect these changes to persist and have long-lasting implications on residents' health and well-being, as e-commerce and flexible work arrangement become commonplaces across different countries. We urge the government and transportation policymakers to pay attention to traffic noise pollution and enhance the optimization of city planning to establish a sustainable transportation system in the post-pandemic world. Our paper has caveats. The extent to which our findings can be generalized to a suburban or urban sprawling metropolitan area requires further research. Due to data limitations, we do not have information on the home conditions, such as closed doors or windows, or air conditioners[5]. We cannot separate the impact of WFH on traffic noise from that of e-commerce on traffic noise, which is warranted for future studies.

[5] However, we do not consider those omitted variables can seriously undermine our estimation, as the public HDB buildings are highly homogeneous in Singapore. For instance, according to data released by the Energy Market Authority of Singapore, as of 2018, the adoption rate of air conditioning among households is as high as 84%.

# Figures and Tables

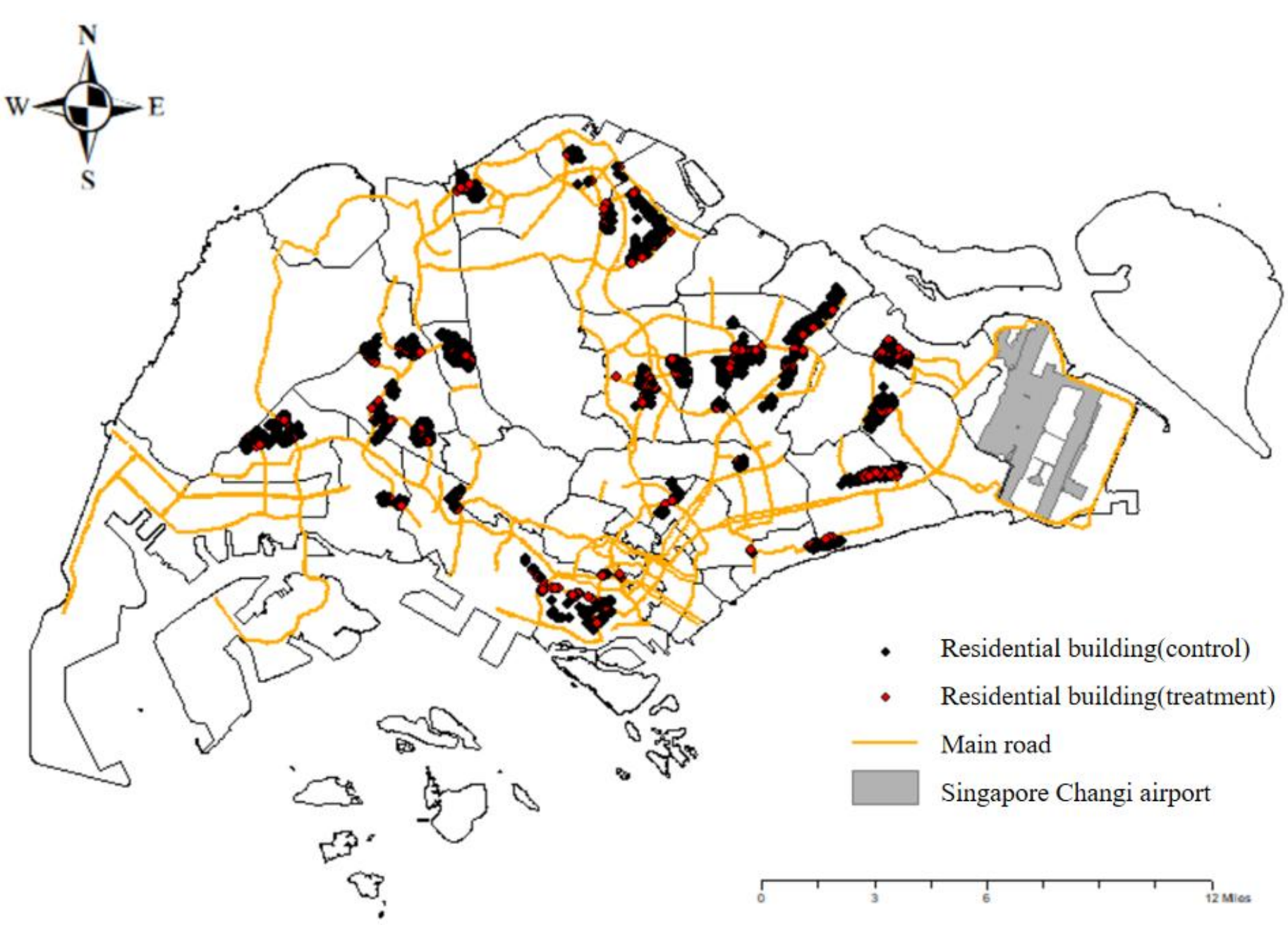

Figure 1. Main road lines and sample distribution in Singapore

*Source: https://www.openhistoricalmap.org/*

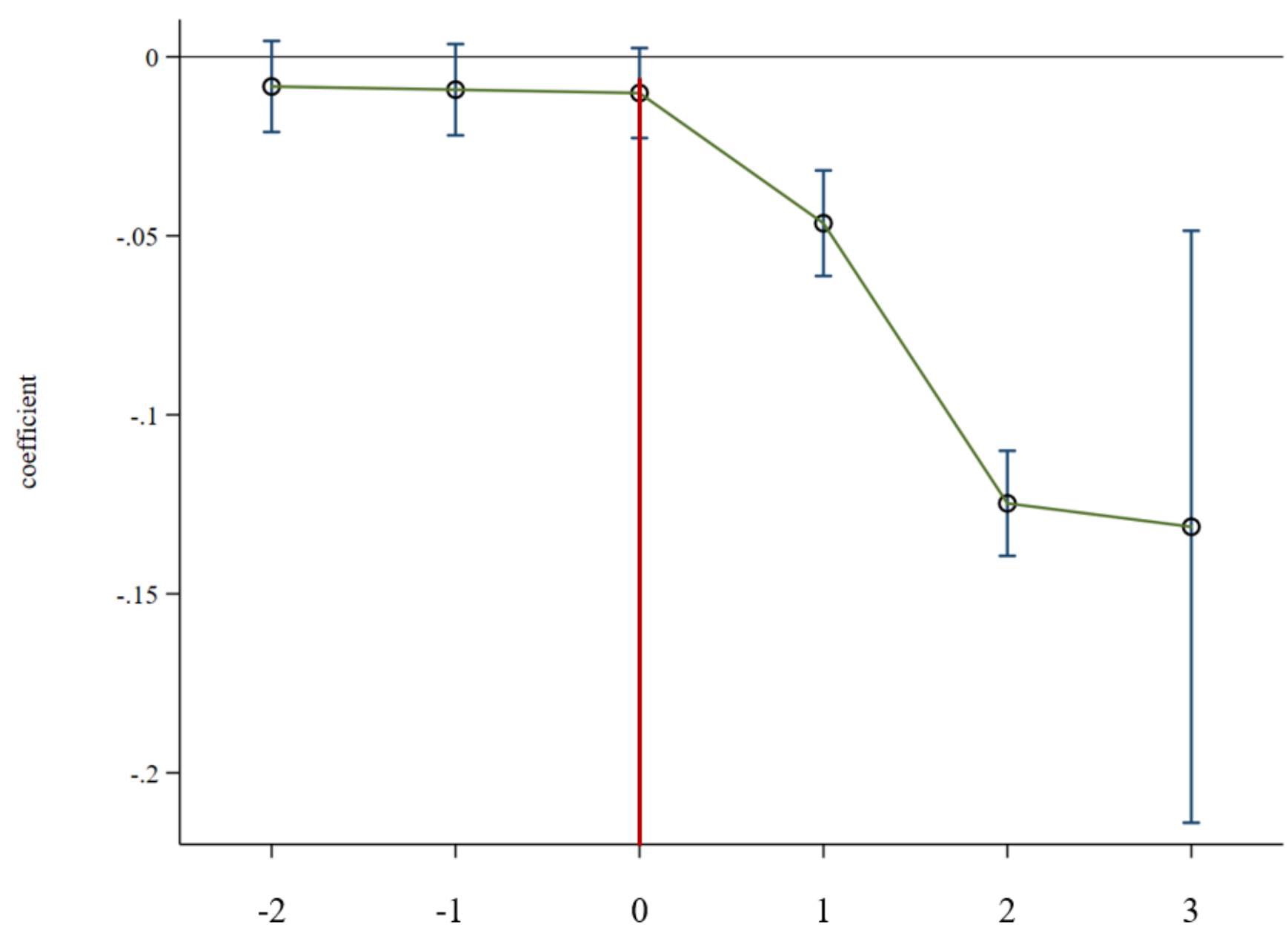

Figure 2. Illustrating the results of the parallel trend test

*Note: The Y-axis presents the coefficients of $Road\ noise \times relative_time_i$. The X-axis represents the six time intervals. Vertical lines indicate the 95% confidence intervals of the coefficients.*

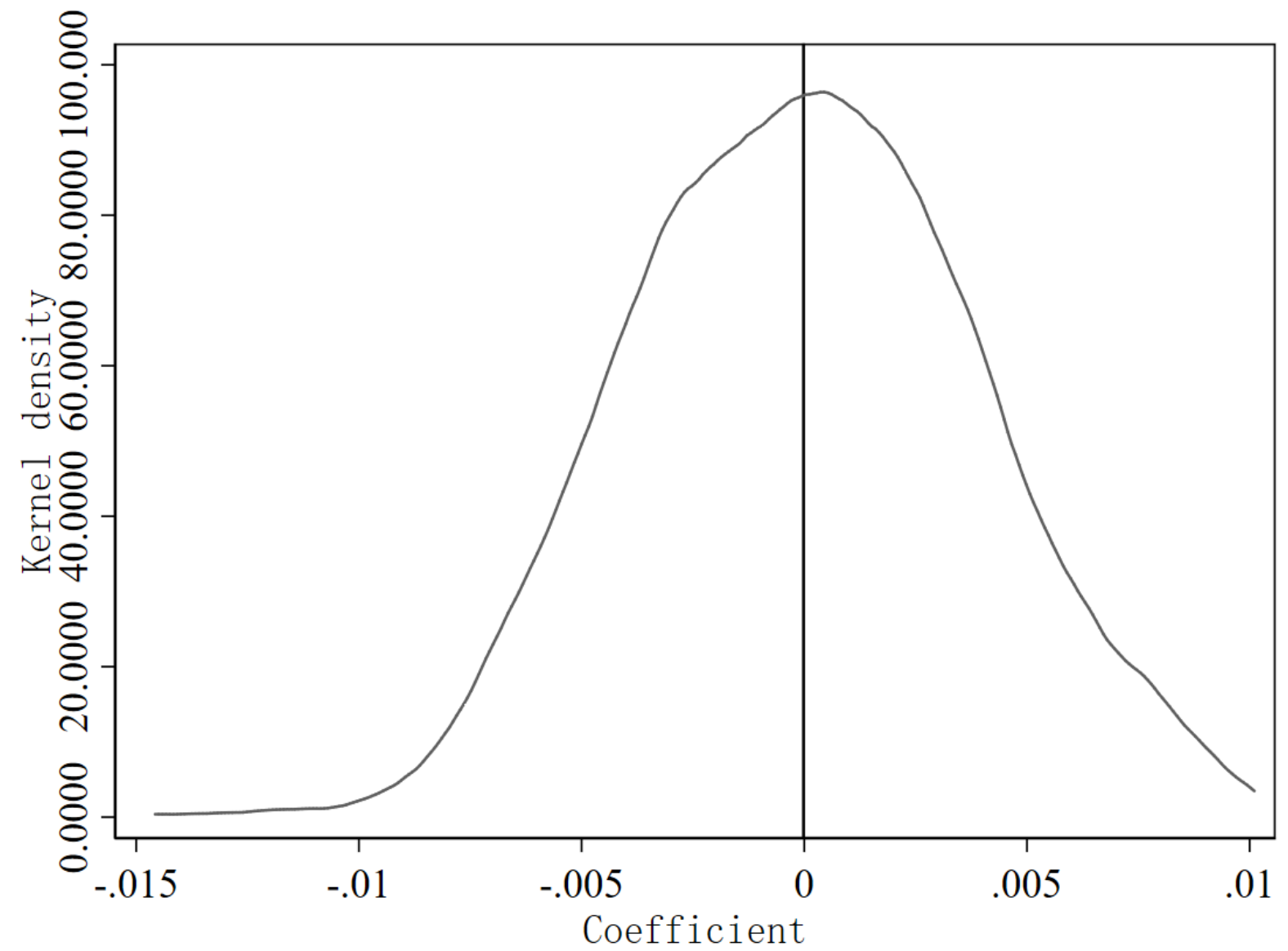

Figure 3.1. Distribution of the 1,000 estimated DID coefficients

*Note: The solid vertical line indicates the mean of the estimated DID coefficients.*

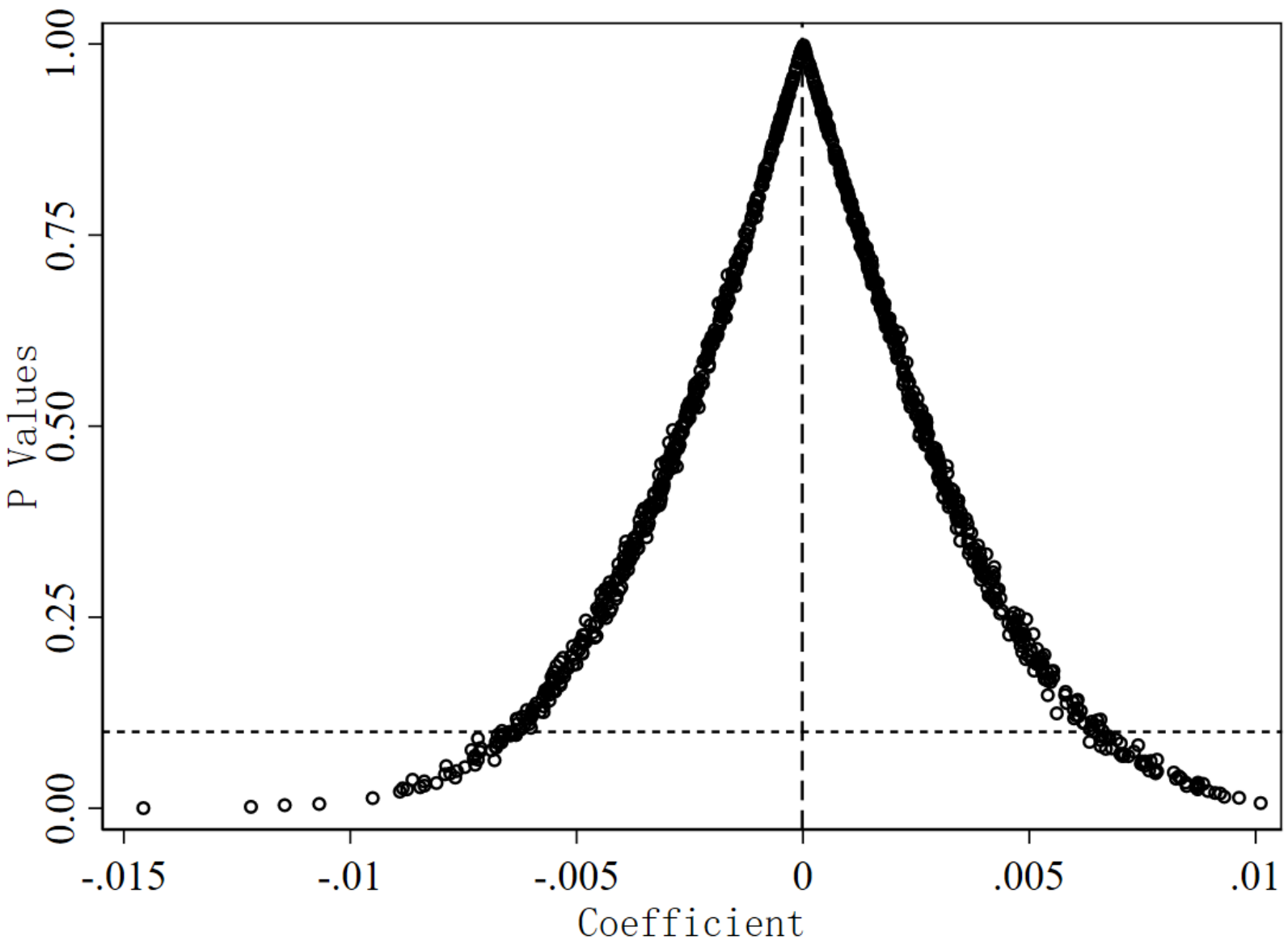

Figure 3.2. Scatter diagram of p-values versus the 1,000 estimated DID coefficients

*Note: The solid vertical line indicates the mean of the estimated DID coefficients, and the dotted horizontal line indicates the 10% p-values.*

Table 1. Summary statistics of key variables

| | Treatment group | | Control group | |
|---|---|---|---|---|
| Number of observations | 13,597 | | 33,384 | |
| | Mean | Std. Dev. | Mean | Std. Dev. |
| *Rent* | 2,033.262 | 404.253 | 2,100.289 | 365.479 |
| *COVID-19* | 0.158 | 0.365 | 0.166 | 0.372 |
| *Size* | 91.668 | 26.359 | 99.264 | 26.536 |
| *Floor* | 7.807 | 4.717 | 8.141 | 4.782 |
| *Age* | 29.577 | 10.882 | 26.181 | 10.588 |
| *Distance to main road* | 61.441 | 22.326 | 272.604 | 114.824 |
| *Distance to expressway* | 1,107.566 | 787.104 | 1241.248 | 845.8549 |
| *Distance to MRT station* | 888.430 | 401.270 | 911.605 | 423.256 |
| *Distance to primary school* | 2,180.934 | 1,717.141 | 2,356.523 | 1,585.479 |
| *Distance to CBD* | 9,553.468 | 4,261.339 | 10,809.640 | 4,178.579 |
| *Distance to bus stop* | 91.670 | 38.326 | 114.233 | 55.749 |

*Source: The 99 Group.*

Table 2. Impacts of road traffic noise on housing rent

| | (1) | (2) | (3) | (4) |
|---|---|---|---|---|
| Variables | Rent | Ln(Rent) | Ln(Rent) | Ln(Rent) |
| | Full sample | Full sample | 1 January 2020–31 December 2020 | 1 January 2021–31 March 2022 |
| *DID* | -171.588*** | -0.083*** | -0.038*** | -0.127*** |
| | (11.304) | (0.005) | (0.006) | (0.006) |
| *COVID-19* | 1,308.853*** | 0.617*** | 0.512*** | 0.632*** |
| | (48.204) | (0.019) | (0.012) | (0.019) |
| *Treatment group* | -8.020* | -0.007** | -0.007*** | -0.007*** |
| | (4.971) | (0.002) | (0.002) | (0.003) |
| *Size* | 1.541*** | 0.285*** | 0.288*** | 0.286*** |
| | (0.560) | (0.012) | (0.013) | (0.012) |
| *Floor* | 2.378*** | 0.005*** | 0.004*** | 0.006*** |
| | (0.366) | (0.001) | (0.001) | (0.001) |
| *Age* | -4.902*** | -0.052*** | -0.052*** | -0.052*** |
| | (0.595) | (0.005) | (0.004) | (0.005) |
| *Ln(Distance to* | -9.657** | -0.004*** | -0.004** | -0.003* |
| *expressway)* | (3.977) | (0.001) | (0.001) | (0.002) |
| *Ln(Distance to MRT* | -96.581*** | -0.040** | -0.042*** | -0.041*** |
| *station)* | (9.198) | (0.000) | (0.004) | (0.004) |
| *Ln(Distance to CBD)* | -138.058*** | -0.069*** | -0.075*** | -0.072*** |
| | (37.993) | (0.069) | (0.018) | (0.018) |
| *Ln(Distance to* | 3.010 | 0.001 | -0.002 | 0.0008 |
| *primary school)* | (7.471) | (0.004) | (0.004) | (0.004) |
| *Ln(Distance to bus* | -0.443 | -0.001 | -0.00001 | -0.0001 |
| *Stop)* | (4.632) | (0.003) | (0.002) | (0.002) |
| *Constant* | 2960.079*** | 6.790*** | 6.865*** | 6.799*** |
| | (389.373) | (0.183) | (0.187) | (0.185) |
| *Year-month fixed* | √ | √ | √ | √ |
| *Housing type-fixed* | √ | √ | √ | √ |
| *District fixed* | √ | √ | √ | √ |
| *Obs* | 46,980 | 46,980 | 43,158 | 43,073 |
| $R^2$ | 0.6094 | 0.6187 | 0.6479 | 0.6332 |

*Note: Values in parentheses are the clustered robust standard error of buildings. All variables are defined in Table 3. *p < 0.1. **p < 0.05 ***p < 0.01. In this research, we conducted regressions using both 'Rent' and 'lnRent' as dependent variables. Due to space limitations, the results for 'Rent' are not reported in the subsequent tables. Those results are available upon request.*

Table 3. Robustness tests

| | (1) | (2) | (3) | (4) |
|---|---|---|---|---|
| | Ln(Rent) | Ln(Rent) | Ln(Rent) | Rent (SGD) |
| | *Distance threshold = 90 m* | *Distance threshold = 110 m* | *Ln(Distance to a main road)* | *Distance to a main road (All sample)* |
| *DID* | -0.078*** | -0.076*** | | |
| | (0.005) | (0.005) | | |
| *COVID-19* Ln(Distance to a main road)* | | | 0.035*** | |
| | | | (0.003) | |
| *COVID-19* | 0.617*** | 0.622*** | 0.417*** | 1217.013*** |
| | (0.021) | (0.022) | (0.027) | (52.967) |
| *Treatment group (90 m)* | -0.010** | | | |
| | (0.003) | | | |
| *Treatment group (110 m)* | | -0.005* | | |
| | | (0.003) | | |
| *Ln(Distance to a main road)* | | | 0.004* | |
| | | | (0.002) | |
| *Distance to a main road* | | | | 0.014 |
| | | | | (0.021) |
| *COVID-19* Distance to a main road* | | | | 0.352*** |
| | | | | (0.034) |
| *Constant* | 7.591*** | 7.581*** | 7.559*** | 771.178*** |
| | (0.205) | (0.206) | (0.212) | (138.817) |
| *X control* | √ | √ | √ | √ |
| *Year-month fixed* | √ | √ | √ | √ |
| *Housing type-fixed* | √ | √ | √ | √ |
| *District fixed* | √ | √ | √ | √ |
| *Obs* | 46,980 | 46,980 | 46,980 | 46,980 |
| $R^2$ | 0.5931 | 0.5925 | 0.5904 | 0.5819 |

*Note: X control indicates a vector of control variables, including size, floor, age; ln(distance to expressway), ln(distance to MRT station), ln(distance to primary school), ln(distance to CBD) and Ln(distance to bus stop). All variables are defined in Table 3a. *p < 0.1. **p < 0.05 ***p < 0.01.*

Table 4. Heterogeneity test

| | (1) | (2) | (3) | (4) | (5) | (6) |
|---|---|---|---|---|---|---|
| | Ln(Rent) | Ln(Rent) | Ln(Rent) | Ln(Rent) | Ln(Rent) | Ln(Rent) |
| | *Less than or equal to the $10^{th}$-floor level* | *More than $10^{th}$-floor level* | *Within 1 km of an MRT station* | *Beyond 1 km of an MRT station* | *Unit size less than or equal to 80 $m^2$* | *Unit size more than 80 $m^2$* |
| *DID* | -0.083*** <br> (0.006) | -0.061*** <br> (0.009) | -0.082*** <br> (0.006) | -0.071*** <br> (0.008) | -0.056*** <br> (0.009) | -0.096*** <br> (0.005) |
| *X control* | √ | √ | √ | √ | √ | √ |
| *Year-month Fixed* | √ | √ | √ | √ | √ | √ |
| *Housing type-Fixed* | √ | √ | √ | √ | √ | √ |
| *District Fixed* | √ | √ | √ | √ | √ | √ |
| *Obs* | 34,518 | 12,462 | 30,720 | 16,260 | 14,831 | 32,149 |
| $R^2$ | 0.5891 | 0.6230 | 0.6024 | 0.6158 | 0.5012 | 0.5926 |

*Note: X control indicates a vector of control variables, including size, floor, age; ln(distance to expressway), ln(distance to MRT station), ln(distance to primary school), ln(distance to CBD) and Ln(distance to bus stop). All variables are defined in Table 3a. *p < 0.1. **p < 0.05 ***p < 0.01.*

# The Price of Quietness:

# How a Pandemic Affects City Dwellers' Response to Road Traffic Noise

# Online Appendix

Table A1. Types of roads in Singapore in 2015

| | Type of road | Road length in kilometres | Road width | Speed limit |
|---|---|---|---|---|
| 1 | Expressway | 164 | Six lanes and above | 90 km/h |
| 2 | Main road | 698 | Four to six lanes | 70 km/h |
| 3 | Collector road | 578 | Four lanes | 50 km/h |
| 4 | Local road | 3,496 | Two lanes | 30–50 km/h |

*Source: Singapore Land Transport Authority Website*

Table A2. Singapore e-commerce revenue and proportion of the population who avoided going to work

| | 2019 | 2020 | 2021 | 2022 |
|---|---|---|---|---|
| E-commerce revenue (USD in billions) | 2.98 | 4.56 | 6.27 | 6.41 |
| Working from home (Share of the population) | Not available | 11%–54% (Peak on 23rd April, first circuit-breaker lockdown) | 17%–41% (Peak on 3rd June, second circuit-breaker semi-lockdown) | 15%–22% (Peak on 12th January) |

*Source: Statista.com*

Table A3. Variable definitions

| Variables | Definition |
|---|---|
| *Rent* | The monthly rent (S$) of a housing unit |
| *COVID-19* | A dummy variable with 1 indicating that a housing unit is transacted after 1 January 2020 and 0 otherwise |
| *Treatment group* | A dummy variable with 1 indicating that a housing unit is exposed to road traffic noise (in the treatment group) and 0 otherwise |
| | The cross term of *Road noise* and *COVID-19* |
| $Group_i$ | A dummy variable with 1 indicating that a housing transaction falls into threshold zone i, i = 1…7, and 0 otherwise (see Section 4.1) |
| *Size* | The size of a housing unit ($m^2$) |
| *Floor* | The floor level of a housing unit in a building |
| *Age* | The age of a housing unit on the transaction date (year) |
| *Year-month fixed* | A set of dummy variables to control time-fixed effect |
| *Housing type-fixed* | A set of dummy variables to control housing type-fixed effect |
| *District fixed* | A set of dummy variables to control location-fixed effect |
| *Distance to main road* | Distance from a residential building to the nearest main road (metres) |
| *Distance to expressway* | Distance from a residential building to the nearest expressway (metres) |
| *Distance to MRT station* | Distance from a residential building to the nearest mass rail transit station (metres) |
| *Distance to primary school* | Distance from a residential building to the nearest primary school (metres) |
| *Distance to central business district (CBD)* | Distance from a residential building to the CBD (metres) |
| *Distance to bus stop* | Distance from a residential building to the nearest bus stop (metres) |

Table A4. Empirical identification of the treatment and control groups

| | (1) | (2) |
|---|---|---|
| Variables | Ln(Rent)<br>Full sample | Ln(Rent)<br>Before the pandemic |
| $group_1$ 0–50 m | -0.024*** | -0.012*** |
| | (0.003) | (0.004) |
| $group_2$ 50–100 m | -0.016*** | -0.004** |
| | (0.003) | (0.001) |
| $group_3$ 100–150 m | 0.003 | 0.003 |
| | (0.003) | (0.003) |
| $group_4$ 150–200 m | 0.004 | 0.003 |
| | (0.003) | (0.003) |
| $group_5$ $200-250$ m | -0.004 | -0.005 |
| | (0.003) | (0.004) |
| $group_6$ 250–300 m | 0.003 | 0.002 |
| | (0.004) | (0.004) |
| *Size* | 0.001*** | 0.001*** |
| | (0.000) | (0.000) |
| *Floor* | 0.001*** | 0.001*** |
| | (0.000) | (0.000) |
| *Age* | -0.002*** | -0.002*** |
| | (0.000) | (0.000) |
| *Ln(Distance to expressway)* | -0.003** | 0.0001 |
| | (0.001) | (0.003) |
| *Ln(Distance to MRT station)* | -0.045*** | -0.045*** |
| | (0.003) | (0.004) |
| *Ln(Distance to CBD)* | -0.051*** | -0.068*** |
| | (0.015) | (0.016) |
| *Ln(Distance to primary school)* | -0.0002 | -0.001 |
| | (0.003) | (0.003) |
| *Ln(Distance to bus stop)* | -0.001 | -0.002 |
| | (0.002) | (0.002) |
| *Constant* | 7.893*** | 8.105** |
| | (0.160) | (0.161) |
| *Year-month fixed* | √ | √ |
| *Housing type-fixed* | √ | √ |
| *District fixed* | √ | √ |
| *Obs* | 46,980 | 39,273 |
| $R^2$ | 0.6189 | 0.6715 |

*Note: Values in parentheses are the clustered robust standard error of buildings. All variables are defined in Table 3. *p < 0.1 **p < 0.05 ***p < 0.01.*

Table A5. Results of the parallel trend test

| Variables | Ln(Rent) |
|---|---|
| $Road\ noise \times relative_time_i$ | Full sample |
| *i=-2* | -0.008 |
| | (0.007) |
| *i=-1* | -0.009 |
| | (0.007) |
| *i=0* | -0.010 |
| | (0.008) |
| *i=1* | -0.046*** |
| | (0.008) |
| *i=2* | -0.124*** |
| | (0.041) |
| *i=3* | -0.131*** |
| | (0.050) |
| *Treatment group* | -0.002*** |
| | (0.000) |
| *constant* | 7.227*** |
| | (0.033) |
| *X control* | √ |
| *Year-month fixed* | √ |
| *Housing type-fixed* | √ |
| *District fixed* | √ |
| *Obs* | 46,981 |
| $R^2$ | 0.3732 |

*Note: X control indicates a vector of control variables, including size, floor, age, ln(distance to expressway), ln(distance to MRT station), ln(distance to primary school), ln(distance to CBD) and ln(distance to bus stop). All variables are defined in Table 3a. *p < 0.1. **p < 0.05 ***p < 0.01.*

Table A6. WTP to noise level

| | Rent (SGD) |
|---|---|
| | *Distance to a main road (All sample)* |
| *Distance to a main road* | 0.014 |
| | (0.021) |
| *COVID-19* Distance to a main road* | $0.352^{***}$ |
| | (0.034) |
| *COVID-19* | $1217.013^{***}$ |
| | (52.967) |
| *Constant* | $771.178^{***}$ |
| | (138.817) |
| *X control* | √ |
| *Year-month fixed* | √ |
| *Housing type-fixed* | √ |
| *District fixed* | √ |
| *Obs* | 46,980 |
| $R^2$ | 0.5819 |

*Note: X control indicates a vector of control variables, including size, floor, age; ln(distance to expressway), ln(distance to MRT station), ln(distance to primary school), ln(distance to CBD) and Ln(distance to bus stop). All variables are defined in Table 3a. *p < 0.1. **p < 0.05 ***p < 0.01.*